\documentclass[12pt]{article}
\usepackage{graphicx}
\title{LOW-MASS PROTON-ANTIPROTON ENHANCEMENT: BELLE AND BES RESULTS, PREMISES OF LEAR AND EXPECTATIONS FROM CLAS
\\}
\author{B.Kerbikov, A.Stavinsky, V.Fedotov\\
 State Research
Center\\Institute of Theoretical and Experimental Physics, \\
Moscow, Russia}
 \date{}
  \newcommand{\be}{\begin{equation}}
\newcommand{\ee}{\end{equation}}
 
\def\fun#1#2{\lower3.6pt\vbox{\baselineskip0pt\lineskip.9pt
\ialign{$\mathsurround=0pt#1\hfil ##\hfil$\crcr#2\crcr\sim\crcr}}}

\newcommand{\vep}{\mbox{\boldmath${\rm p}$}}

\begin{document}
\maketitle

\begin{abstract}
We present a simple explanation for the recently observed
near-threshold proton-antiproton enhancement. It is described by a
set of low-energy parameters deduced from the analysis of $N\bar
N$  experiments at LEAR. We predict a related effect in
photoproduction reaction under study by CLAS collaboration.
\end{abstract}

\section{Introduction}

Low-mass baryon-antibaryon enhancement has recently been observed
in the decays $B^+\to K^+p\bar p$ \cite{1}, $\bar B^0 \to D^0p\bar
p $ \cite{2} and $J/\psi \to \gamma p\bar p$ \cite{3}. On the
theoretical side this discovery has been discussed by several
authors   \cite{4,5,6}. In \cite{4} the near-threshold effect has
been considered either as a gluonic state or as a result of the
quark fragmentation process. In \cite{5} it has been attributed to
the Breit colorspin interaction while in \cite{6} it has been
regarded as caused by peripheral  one-pion-exchange potential.

The history of bound-states and resonances in $N\bar N$ system
("baryonium") started before the discovery of the antiproton and
is full of controversial experimental findings and theoretical
claims. The interested reader is addressed to review papers
\cite{7}-\cite{11}. At the end of eighties an ample supply of
diverse information of low-energy $N\bar N$ interaction has been
obtained mainly due to the commissioning of the Low Energy
Antiproton Ring (LEAR) at CERN. In the situation when theoretical
models were at variance with each other and unable to explain the
whole set of the observed phenomena the semi-phenomenological
effective range analysis \cite{12}-\cite{14} turned out to be a
coherent approach to low-energy $N\bar N$ interaction. In this
approach the lack of the dynamical picture is traded for the
possibility to reconcile within a unique scheme different pieces
of information on $N\bar N$ interaction.

As we shall see the effective range analysis \cite{12}-\cite{14}
based on the old LEAR data enables to explain Belle \cite{1,2} and
BES \cite{3} results. In particular the observation by BES of a
strong effect in the decay $J/\psi\to \gamma p \bar p$ and the
absence of a similar structure in $J/\psi \to \pi^0 p\bar p$
perfectly fits into the solution for the low-energy parameters
obtained 15 years ago \cite{12}-\cite{14}.  We shall return to
this point below. An immediate consequence of the proposed scheme
is the prediction of a related effect in the reaction $\gamma p
\to pp\bar p $. This process can be investigated by CLAS
collaboration at Jefferson Lab. Unlike $B$- and $J/\psi$ decays
the photoproduction reaction has not been discussed from the
theoretical side. Therefore we shall choose it as a starting point
in order to introduce the effective range formalism which also
applies to Belle and BES data.

The double differential cross section for the reaction $\gamma p
\to pp\bar p$ is given by the well-known Chew-Low expression
\be
\frac{d^2\sigma}{ds_2 dt_1} = \frac{1}{2^{12}\pi^4}
\frac{\lambda^{1/2} (s_2, m^2, m^2)}{m^2k^2s_2} \int
d\Omega^*_{23}|T(s_2, t_1,\Omega^*_{23})|^2,\label{1} \ee where
$m$ is the nucleon mass, $k$ is the energy of $\gamma$,
\be
s_2=(p_2+p_3)^2,~~t_1= (k-p_1)^2,\label{2}\ee with $p_2$ and $p_3$
being the   4-momenta of $p$ and $\bar p$ forming the low-mass
pair, $p_1$ being the 4-momentum of the remaining proton,
$\lambda(x,y,z) = (x-y-z)^2 -4yz$ is the standard kinematical
function. The angle $\Omega^*_{23}= (\cos \theta, \varphi) $ is
defined in the CM system of the particles  (2,3). The variable
$s_2=m^2_{23}$ is the square of the invariant mass of the (2,3)
pair. It follows from (\ref{1}) that the angular distribution of
particles 2 and 3 ( the low-mass $p\bar p$ pair) is isotropic in
their CM system provided $T(s_2, t_1, \Omega^*_{23})$ is
independent of $\Omega^*_{23}$ which is the case for $^1S_0$ and
$^3P_0$ states. Equation (\ref{1}) yields the following invariant
mass distribution \be \frac{d\sigma}{dm_{23}} =
\frac{1}{2^{10}\pi^4} \frac{|\vep_1|}{km^2}
({m^2_{23}-4m^2})^{1/2}\int d(\cos\beta) \int d\Omega^*_{23}
|T|^2,\label{3}\ee where $\beta $ is the angle between $\vep_1$
and   the direction of  the incident $\gamma$ ( we remind that the
index 1 is attributed to the proton which remains outside the
correlated  $p\bar p$ pair).

In the spirit of the Migdal-Watson FSI theory we single out from
the matrix element $T$ a factor responsible for the low energy $p
\bar p$ interaction. Within Migdal-Watson approach one has
\be
|T|^2 = |T^{(0)}|^2/|f(-q)|^2 = D(q) |T^{(0)}|^2,\label{4}\ee
where $q$ is the $p\bar p$ CM momenta, $q^2=\frac14
(m_{23}^2-4m^2)$, $f(-q)$ is the Jost function corresponding to
the $p\bar p$ interaction at low energy, $D(q) = |f(-q)|^{-2}$ is
called the enhancement factor. As it was already mentioned the
dynamics of $N\bar N$ interaction  is much more complicated than
that of $NN$ \cite{7} -\cite{11}. Annihilation  dominates at short
and possibly intermediate distances, $\omega$- and other odd
$G$-parity exchanges lead to a strong attraction to be added to
 the one-pion exchange in the outer region \cite{15}. Therefore the
approximation of $f(-q)$ by a Born term from one-pion exchange
\cite{6} misses the essential features of $N\bar N$ dynamics. On
the other hand effective range solution for the low-energy $N\bar
N$ amplitude describes the whole set of data with a fair accuracy
\cite{12}-\cite{14}.

Ignoring for the moment complications due to annihilation, Coulomb
interaction and spin-isospin structure we may write the following
expression for Jost function in the scattering length
approximation
 \be
f(-q) \simeq A(q) (1-iqa),\label{5}\ee where $a$ is the scattering
length and  $A(q)$ is a well defined smooth function with a
property $A(q\to 0)\to 1$. A similar expression for $p\bar p$
system is  not as simple for  three reasons:

(i) $p\bar p$ system is a combination of the two isospin states
with $I=0,1$; isospin invariance is violated by the mass
difference of proton and neutron;

(ii) a powerful annihilation results in the complexity of the
scattering lengths;

(iii) the Coulomb attraction acts in the $p\bar p$ system.

One should also keep in mind that  there are two spin states with
 $S=0,1$.
   In the absence of polarization experiments only
 spin-averages quantities were extracted from the experimental
 data \cite{12}-\cite{14} (see also \cite{16}).

 The enhancement factor in which all three points (i) -(iii)
 inherent for the $p\bar p$ system are accounted for has the
 following form \cite{12}-\cite{14}
 \be
 D(q) =\frac{c^2(q)}{|1-is (\tilde q
 +i\Delta  +l)-(\tilde q + i\Delta) lr|^2},\label{6}\ee
 where
 \be
 c^2(q) = \frac{2\pi}{qa_B}\left [ 1-\exp \left( -
 \frac{2\pi}{qa_B}\right) \right]^{-1}\label{7}\ee
 is the Sakharov Coulomb attraction factor \cite{17} with $a_B =2/\alpha m =57.6$
 fm being the Bohr radius of the  $p\bar p$ atom,
 \be
 s=\frac12 (a_0+a_1), ~~ r=a_0a_1,\label{8}\ee
 where $a_I, I=0,1$ are $N\bar N$  $S$ -wave scattering  lengths
 with isospin $I$, \be
 \tilde q = c^2 (q) q +  \frac{2i}{a_B} h(qa_B),~~ h(z) =\ln z
 +Re\psi \left (1+\frac{i}{z}\right),\label{9}\ee
with  $\psi (z) =d/dz\ln \Gamma(z).$ The  quantity $l$ is the
 momentum in the $n\bar n$ channel,
\be
 l^2=q^2-m_n\delta,~~ \delta = 2(m_n-m),\label{10}\ee
 (recall that $m$ is the proton mass).

 The point $q=({m\delta})^{1/2} \simeq 49$ MeV/c corresponds to the
 $n\bar n$ threshold; below this point $n\bar n$ momentum $l$
 becomes imaginary, $l=i(m\delta-q^2)^{1/2}.$
Parameter $\Delta $ is the Schwinger correction to the scattering
length \cite{18}, $\Delta \simeq - 0.08$ fm$^{-1}$ \cite{12}-\cite{14}.

The enhancement factor (\ref{6}) does not include the effective
range term and the contribution from nonzero orbital momenta.
According to \cite{12}-\cite{14} the effective range term is of
minor importance up to $q\simeq 150$ MeV/c i.e. to
$Q=m_{23}-2m\simeq 20$ MeV. We  note  in passing that for a
multichannel system the effective range may be negative or  even
complex \cite{19}. The $P$-wave  comes into play  a little earlier
especially as far as the total cross section is concerned.
Inclusion of  $P$-wave brings about two problems. First, the
$P$-wave  scattering  length (it has a dimension of fm$^3$) is
very sensitive to the fitting procedure (see \cite{12}-\cite{14})
and the present set of the experimental  data do not warrant a
stable solution for the $l=1$ amplitude. The second problem is the
following. With $P$-wave included equations become rather
cumbersome and the number of parameters increases substantially.
This may cause unnecessary doubts in the  reliability of the
proposed approach. We plan to consider effective range terms and
$l>0$ amplitudes in the next  publication. This would allow to
analyze the $p\bar p$ correlation function in a wider energy
range.

The most remarkable result of all fits \cite{12}-\cite{14} for low
energy $N\bar N$ parameters is a  clear  dominance of the $S$-wave
with $I=0$ over that with $I=1$. The corresponding scattering
lengths read
\be
a_0 = (-1.2+ i0.9) {\rm ~fm},~~ a_1 = (-0.1+ i0.4) {\rm
~fm},\label{11}\ee
\be
a_0 = (-1.1+ i0.4) {\rm ~fm},~~ a_1 = (0.3+ i0.8) {\rm
~fm},\label{12}\ee where (\ref{11}) and (\ref{12}) are
respectively the results of \cite{13} and \cite{14}. The sign
convention in \cite{12}-\cite{14} was $ k \cot \delta = +1/a$. We
see that the absolute value of  $|Rea_0|$  is much larger than
that of  $|Re a_1 |$. This is completely in line with the
observation by BES of a strong low-mass effect in the decay
process $J/\psi\to \gamma p \bar p$ and the absence of a similar
structure in $J/\psi \to \pi^0 p\bar p $ \cite{3}.

The negative sign of $Rea_0$ may be interpreted  as an indication
that the potential in this channel is either repulsive or on the
contrary attractive and  strong enough to produce a bound state
somewhere below  the threshold. However such a simple
interpretation is not obvious in strong annihilation regime. This
problem is beyond the scope of the present work.

Coming back to Eq.(\ref{3}) for the effective mass distribution we
note that due to Coulomb attraction it has a tiny jump at
threshold since $q c^2 (q) \to 2\pi/a_B$ at  $q\to 0$ and it has
cusp at $q= (m\delta)^{1/2}$ due to  the opening of the $n\bar n$
channel.

As it is clear  from Eqs. (\ref{3}-\ref{4}) the near threshold
region is dominated by the interplay of the phase-space factor
$(m^2_{23}-4m^2)^{1/2}$ and the enhancement factor $D(q)$. The
first one is  trivial and plays a role of a universal background.

The two solutions given by (\ref{11}) and (\ref{12}) are somewhat
different. This is due to several reasons described in
\cite{12}-\cite{14}. The main source of ambiguity is the lack of
the experimental data at very low energies. Therefore, Belle, BES
and future CLAS data which we have interpreted in the framework of
the effective range solution may in their turn help to update the
solution itself.

In Fig.1 we plot the enhancement  factor $D$ as a function of
\be
Q=m_{23} -2m = 2m \left\{ \left(1+\frac{q^2}{m^2}\right)^{1/2}
-1\right\}.\label{13}\ee

The solid curve corresponds to the solution (\ref{11}) while the dashed one
to the solution (\ref{12}).

The authors are indebted to L.N.Bogdanova, K.Michailov and
Yu.A.Simonov for useful discussions and clarifying remarks.

\newpage

\begin{picture}(0,530)
\put(-100,0){\includegraphics[width=1.4\textwidth]{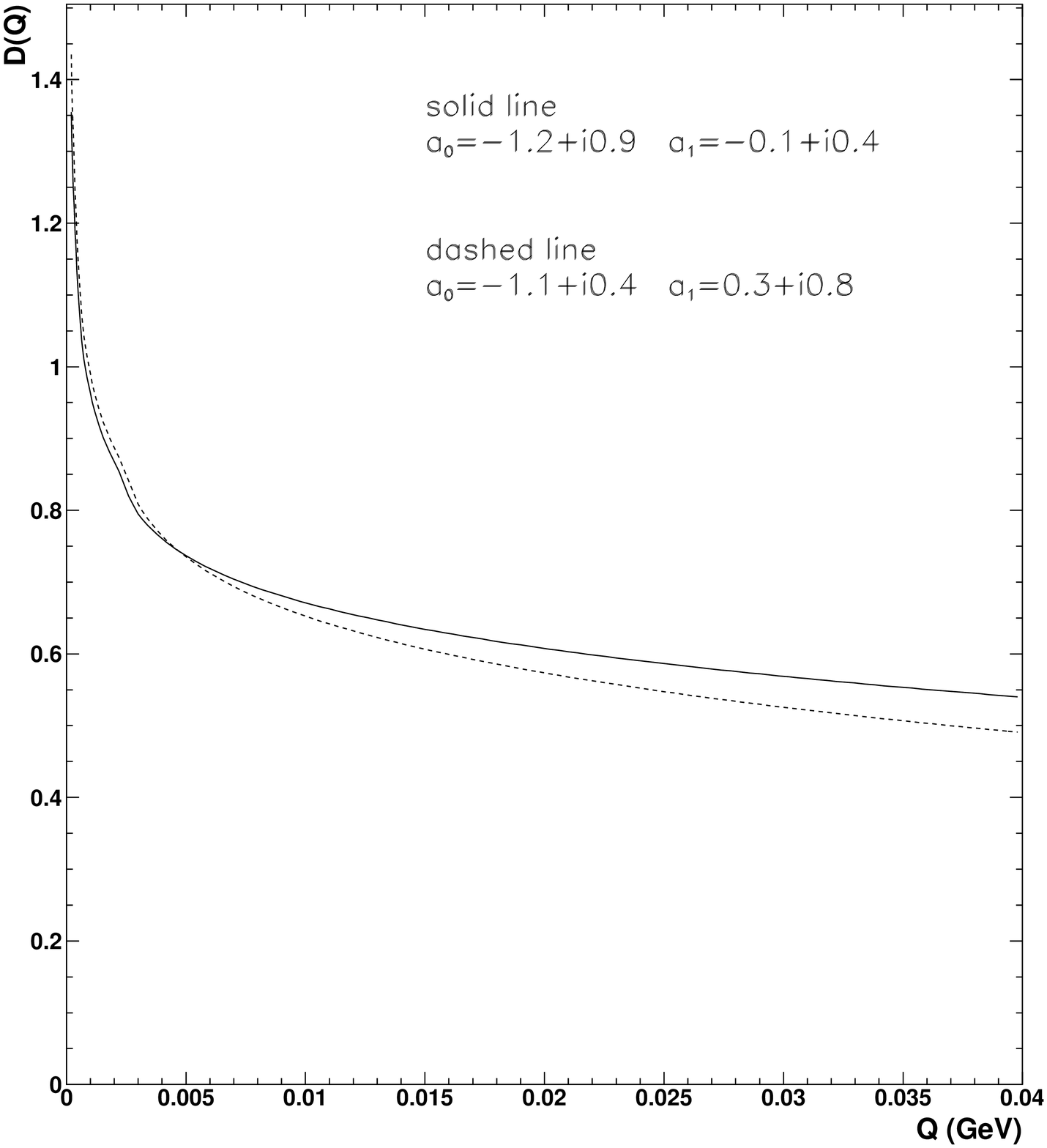}}

\put(-50,-20){Fig.1 The near-threshold  behaviour of the
enhabcement function $D(Q)$, $Q= m_{23}-2m$.}
\put(-50,-35){See the text for the description of the curves.}

\end{picture}

\end{document}